\documentclass[apj]{emulateapj}

\usepackage{natbib}
\newcommand{\thisstar}{HR\,3549\,}

\slugcomment{Mawet et al.~2015, ApJ, in press}

\shorttitle{Discovery of a low-mass companion around \thisstar \footnotemark[1]}
\shortauthors{Mawet et al.}

\begin{document}


\title{Discovery of a low-mass companion around \thisstar}


\author{D. Mawet\altaffilmark{1}, T. David, M. Bottom} 
\affil{Department of Astronomy, California Institute of Technology, 1200 E. California Blvd, MC 249-17, Pasadena, CA 91125 USA}

\author{S. Hinkley}
\affil{Department of Physics and Astronomy, University of Exeter, Physics Building, Stocker Road, Exeter EX4 4QL}

\author{K. Stapelfeldt, D. Padgett}
\affil{NASA Goddard Space Flight Center, 8800 Greenbelt Rd, Greenbelt, MD 20771 USA}

\author{B. Mennesson, E. Serabyn, F. Morales, J. Kuhn}
\affil{Jet Propulsion Laboratory, California Institute of Technology, 4800 Oak Grove Drive, Pasadena, CA 91109}


\altaffiltext{1}{Jet Propulsion Laboratory, California Institute of Technology, 4800 Oak Grove Drive, Pasadena, CA 91109}
\email{dmawet@astro.caltech.edu}

\footnotetext[1]{Based on observations made with ESO Telescopes at the La Silla Paranal Observatory under programs: 090.C-0486A and 094.C-0406A.}


\begin{abstract}
We report the discovery of a low-mass companion to \thisstar, an A0V star surrounded by a debris disk with a warm excess detected by WISE at 22 $\mu$m ($10\sigma$ significance). We imaged \thisstar B in the L-band with NAOS-CONICA, the adaptive optics infrared camera of the Very Large Telescope, in January 2013 and confirmed its common proper motion in January 2015. The companion is at a projected separation of $\simeq 80$ AU and position angle of $\simeq 157^\circ$, so it is orbiting well beyond the warm disk inner edge of $r > 10$ AU. Our age estimate for this system corresponds to a companion mass in the range 15-80 $M_J$, spanning the brown dwarf regime, and so \thisstar B is another recent addition to the growing list of brown dwarf desert objects with extreme mass ratios. The simultaneous presence of a warm disk and a brown dwarf around \thisstar provides interesting empirical constraints on models of the formation of substellar companions.
\end{abstract}





\section{Introduction}

Although high contrast imaging of self-luminous exoplanets or brown dwarfs around young stars is already difficult because of the angular resolution and contrast required, a further complication is that the interpretation of the nature of any companions found in these images is made difficult because it relies upon theoretical models of formation and evolution that depends on a good knowledge of the host star characteristics. While instruments, observing strategies, and post-processing techniques have undeniably improved over the past decade, accurate stellar age determination is becoming the most pressing challenge of characterization. Because they are a robust yet imprecise sign of youth, it is no coincidence that the majority of exoplanets imaged by adaptive optics are found in systems with debris disks: HR8799 \citep{Marois2008}, $\beta$ Pictoris \citep{Lagrange2010}, HD95086 \citep{Rameau2013}, Fomalhaut \citep{Kalas2008}, GJ758 \citep{Thalmann2009}, and HD106906 \citep{Bailey2014}. We note that the GJ504 system is, out of all bona fide directly imaged exoplanets, currently the only one without confirmed infrared excess, but is also the subject of an on-going controversy about its age. \citet{Fuhrmann2015} indeed suggests that the GJ504 system is actually of Solar age, and that GJ504b is thus likely not a $\sim 4$ Jupiter mass planet, but a 4.5 Gyr old Brown Dwarf. 

Debris disks are the signposts of planetary systems. Collisions among asteroidal and cometary parent bodies maintain the observed dust population against losses to radiative forces. Since dust production is enhanced by gravitational stirring, debris disks systems are natural targets for giant planet imaging searches. It has been established both theoretically and observationally that warm dust is transient. Its presence thus serves as a marker for a young or dynamically active planetesimal belt, and clearly indicates that the host star possesses some kind of planetary system. Like self-luminous exoplanets, the brightness of a debris disk decays with time. To first order, the magnitude of the warm excess serves as a chronometer: 22/24 $\mu$m flux densities larger than 1.2$\times$ the stellar photosphere are almost always found in sources younger than 1 Gyr, and more typically, in stars with ages less than a few hundred Myr. This theoretical expectation has been borne out through both observational work and computational modelling. \citet{2005ApJ...620.1010R} have shown that warm excess, as observed at 24 $\mu$m, is a steeply declining function of stellar age in a sample of 266 A stars.  Models of the dynamical evolution of planetesimal swarms confirm this behaviour for A stars \citep{2007ApJ...663..365W}, punctuated by transient spikes of higher dust content following major planetesimal collisions.  

Here we report the detection of a bound low-mass companion to dusty host star \thisstar. This discovery, reminiscent of $\kappa$ Andromedae \citep {Carson13, 2013ApJ...779..153H}, adds to the collection of extreme mass ratios sub-stellar companions filling the brown dwarf desert.

\begin{deluxetable}{ll}
\tabletypesize{\scriptsize}

\tablecaption{Properties of \thisstar \label{tbl-1}}
\tablewidth{0pt}
\tablehead{\colhead{Properties} & \colhead{value}}
\startdata
Identifiers &  \thisstar, HD 76346, HIP 43620  \\
Coord. (hms, dms) & 08 53 03.77832 -56 38 58.1462 \\
Galactic coord. (deg) & 274.35 -07.66 \\
Spectral type & A0V \\
Distance (pc) & $92.5 \pm 2.5$ \\
V mag  & 6.01  \\ 
L mag & $6.04 \pm 0.05$ \\
$\mu_\alpha$ (mas/yr) &$-22.81 \pm 0.26$ \\ 
$\mu_\delta * \cos{\delta}$ (mas/yr) &$36.54 \pm 0.26$ \\
RV (km/s) & $23.90 \pm 2.2$ \\
$[3.6]-[22]\mu$m & $0.56 \pm 0.06$\\

$T_\mathrm{eff}$  (K)                                   & 10207 $\pm$ 347 \\
$\log{(g)}$ (cgs)                               & 4.20 $\pm$ 0.14    \\
$v\sin{i}$ (km/s)                           & 212 \citep{2005yCat.3244....0G} \\
Median Bayesian age (Myr)   & 230 \\
68\% age CI (Myr)           & 120-360 \\
95\% age CI (Myr)           & 10-390 \\
2D interpolated age (Myr)   & 200$^{+20}_{-160}$ \\
Median Bayesian mass (M$_\odot$) & 2.32 \\
68\% mass CI (M$_\odot$)         & 2.2-2.5 \\
95\% mass CI (M$_\odot$)         & 2.1-2.6 \\
\enddata
\end{deluxetable}

\section{Fundamental parameters of \thisstar}
\label{sec:section2}

\thisstar (HD 76346, HIP 43620) is a main sequence A0 star of visual magnitude 6 (Table~\ref{tbl-1}). The parallax measured by Hipparcos \citep{2007A&A...474..653V} is $10.82\pm 0.27$ mas, corresponding to a distance of $\simeq 92.5$ pc. The WISE satellite measured W1-W4 = 0.56 $\pm 0.06$ mag of excess at 22 $\mu$m, and no excess at 12 $\mu$m (\citet{2013wise.rept....1C}, data from AllWISE cryogenic sky survey). For an A0 photosphere, this excess corresponds to a minimum $L_{dust}/L_{star}$ of $10^{-4}$, similar to planet-bearing HR 8799. This value is for the case of a blackbody that peaks at 22 $\mu$m. The presence of any emission extended over a range of temperatures would add to this value.

We used the Bayesian Analysis for Nearby Young AssociatioNs II (BANYAN II) online tool to determine the membership probability of \thisstar to nearby young kinematic groups \citep{2013ApJ...762...88M, 2014ApJ...783..121G}. This tool is based on a comparison of Galactic position (XYZ) and space velocity (UVW) to well-defined moving groups closer than 100 pc and younger than 200 Myr. No clear association can be found with any known nearby young moving group, so \thisstar is likely a field A0 star.

Our direct interpolation in $\log{(T_\mathrm{eff})}-\log{(g)}$ space relative to PARSEC v1.1 evolutionary models \citep{2012MNRAS.427..127B} yields 200 Myr, with a 40-220 Myr 68\% confidence interval, determined from a distribution of 50,000 interpolated ages reached through Monte Carlo propagation of the associated errors in $\log{(T_\mathrm{eff})}-\log{(g)}$. However, as pointed out in \citet{David2015}, direct interpolation does not account for the nonlinear mapping of time onto the H-R diagram, nor the non-uniform distribution of stellar masses observed in the galaxy, and can lead to biases. 
 
\begin{figure*}
\includegraphics[angle=0,width=10.75cm]{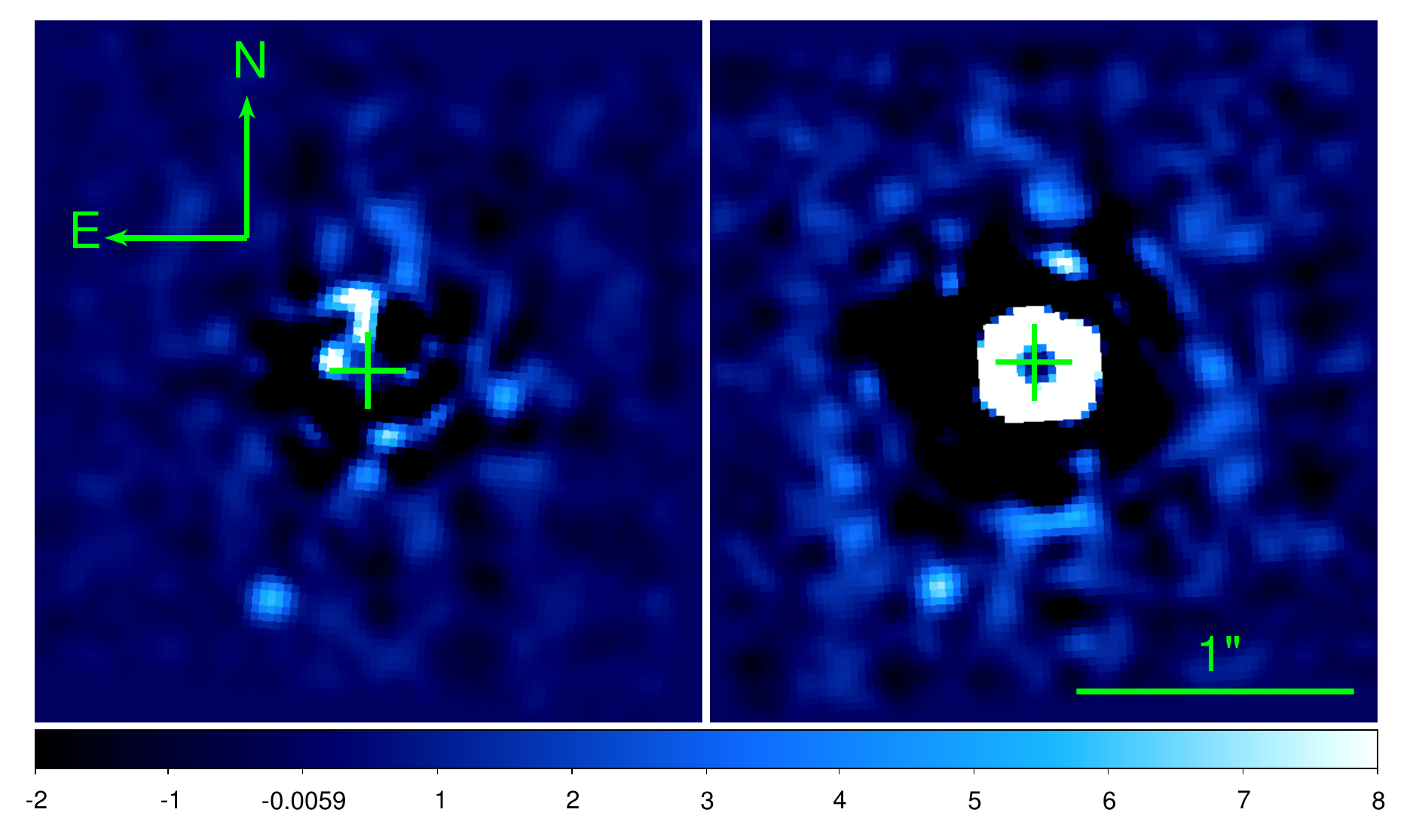}
\includegraphics[width=10cm]{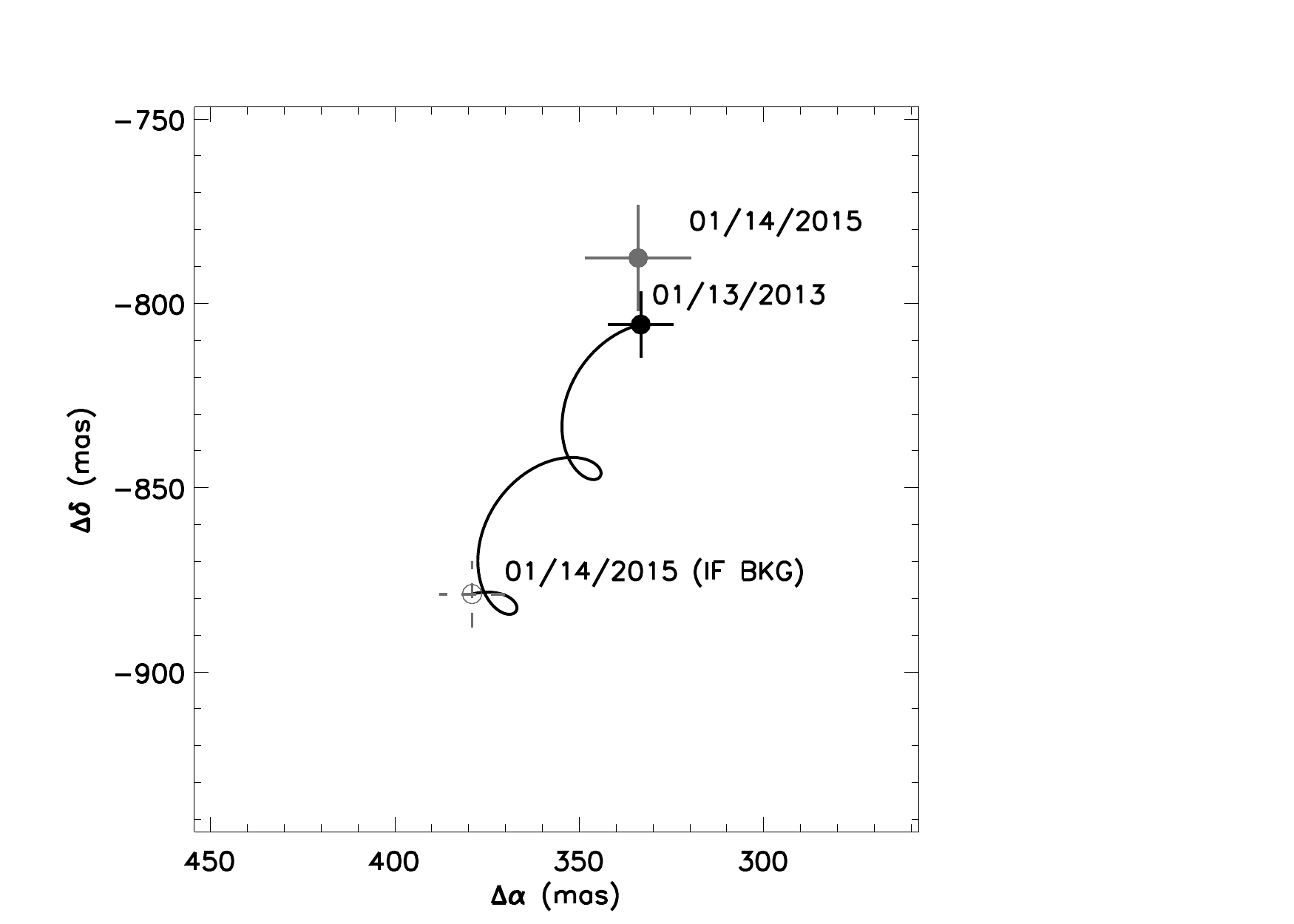}
\caption{Left: discovery image of \thisstar B, taken in January 2013, and confirmation image in January 2015. Right: Common proper motion analysis of \thisstar B. The black dot shows the relative position of the candidate companion in 2013. The empty circle shows the relative position of the point source in 2015 if it were a background object, accounting for parallactic motion (plain curve). The gray dot is the measured position of the bound companion in 2015. \label{fig1}}
\end{figure*}

\section{Observations and data reduction}

We observed \thisstar with NAOS-CONICA at the Very Large Telescope as part of program 090.C-0486(A), ``L-band adaptive optics imaging of exoplanets around a sample of dusty A stars recently discovered by WISE'' (PI: Mawet) on 2013-01-13. We used the Lp-band filter. L-band (centered around 3.8 $\mu$m) is a compelling and competitive filter for ground-based planet surveys. This wavelength range offers significant advantages compared to shorter wavelengths: (i) The L-band contrast of planetary-mass companions with respect to their host stars is more favorable than in the H and K bands \citep{Baraffe2003} so that lower-mass objects can be probed; and (ii) longer wavelengths lead to better image quality and a more stable point-spread function (PSF), with Strehl ratios well above 70\%. These advantages outweigh the increased sky background in the thermal infrared and the loss in resolution. Finally, we note that background star contamination probability rates will be near-zero at small separations in L-band, so minimal follow-up time will be needed to confirm candidates.

The data were acquired as a sequence of 10 exposures on a 5-point dithering pattern with offsets of about 6 arcseconds each. Each exposure was the average of 100 frames with 0.25 s integration time, making for a total open shutter time of 250 s. We used the pupil tracking mode where the instrument co-rotates with the telescope pupil to fix diffraction and speckles to the detector reference frame, allowing the sky to counter-rotate with the parallactic angle, effectively enabling angular differential imaging \cite[ADI][]{Marois2006}. ADI requires sufficient sky rotation to avoid self-subtraction of the companion signal at small angles, often leading to long sequences. Our strategy to overcome this limitation of ADI has been to limit the sequence duration and rely on our uniform target sample to build a library of reference PSFs, and perform reference star differential imaging (RDI). Indeed, if the sample is uniform in brightness, covers a reasonable range of observing conditions and is spread uniformly across the sky, correlated speckle patterns can be retrieved in the library of PSFs. 

We reduced the data by subtracting a background made out of median-combined dithered frames, dividing by a flat field and interpolating for bad pixels and other cosmetics. The reduced images were then processed by two independent speckle calibration packages based on principal component analysis \citep{Soummer2012}, and the library of reference PSF for the speckle calibration.

\section{Discovery of a candidate low-mass companion to \thisstar}

The 2013-01-13 data set showed a point source at a separation of $\simeq 0\farcs 9$, and position angle (PA) of $\simeq 157^\circ$. The detection is unambiguous with a signal-to-noise ratio (SNR) of $\simeq 6$ (see Fig.~\ref{fig1}, left). The TRILEGAL starcount model \citep{Girardi2012} estimates a probability of $7\times 10^{-4}$ that it is an unrelated background object. While this number must be interpreted with caution, it is significant enough to warrant follow-up observations. 

We acquired the second epoch data set as part of program 094.C-0406(A), ``L-band adaptive optics imaging of exoplanets around a sample of dusty stars recently discovered by WISE. Part III: candidate follow-ups'' (PI: Mawet), on 2015-01-14, following a similar strategy as for the discovery epoch. The candidate was detected at roughly the same location though with a lower SNR ($\simeq 3$, see Fig.~\ref{fig1}, right). The lower SNR appears to be mostly due to the characteristics of the Aladdin2 detector installed in CONICA in January 2015 to replace the faulty Aladdin3 detector, and the additional thermal background induced by the different shielding characteristics of this detector configuration.


\section{Astrometry and common proper motion analysis}

We performed astrometric and photometric measurement of both epochs using forward modelling of the off-axis companion point spread function with a Monte-Carlo Markov chain (MCMC) sampler in the $\alpha, \delta, flux$ space. The MCMC sampler model also takes into account variable error bars, meaning that systematic under-reporting of errors should not affect the final result. 

The final error bars on the astrometry were then conservatively set to the quadratic sum of the MCMC-derived error bars, which can still be biased by underlying speckle noise, and the empirically derived influence of SNR on $\alpha, \delta, flux$. Indeed, astrometric precision \citep{Guyon2012} is proportional to $\mathrm{FWHM} / (2\times \mathrm{SNR})$ in the speckle-noise dominated regime, where FWHM is the full width at half maximum of the resolution element $\lambda/D$, with observing wavelength $\lambda$ and telescope diameter $D$. 

As far as the astrometric calibration of CONICA is concerned, we used the $27.1 \pm 0.04$ mas/pixel plate scale and $-0.45 \pm 0.09$ degree true North offset of \citet{Absil2013} for the 2014 epoch, and the $26.99 \pm 0.02$ mas/pixel plate scale and $+0.31 \pm 0.02$ degree true North offset for the 2015 epoch, measured after the NACO recommissioning with the Aladdin2 chip on UT1 (personal communication from NACO instrument scientist Julien Girard, European Southern Observatory). These additional systematic offsets and corresponding error bars were folded into both epoch astrometric positions and associated error bars (Table~\ref{tbl-2}) to perform the common proper motion analysis (CPM) shown in Fig.~\ref{fig1} (right). Following the CPM analysis, we determined that the probability of the discovered candidate to be a background object is $2\times 10^{-8}$. 

\section{Age of \thisstar}

\begin{figure*}
\centering
\includegraphics[width=0.9\textwidth]{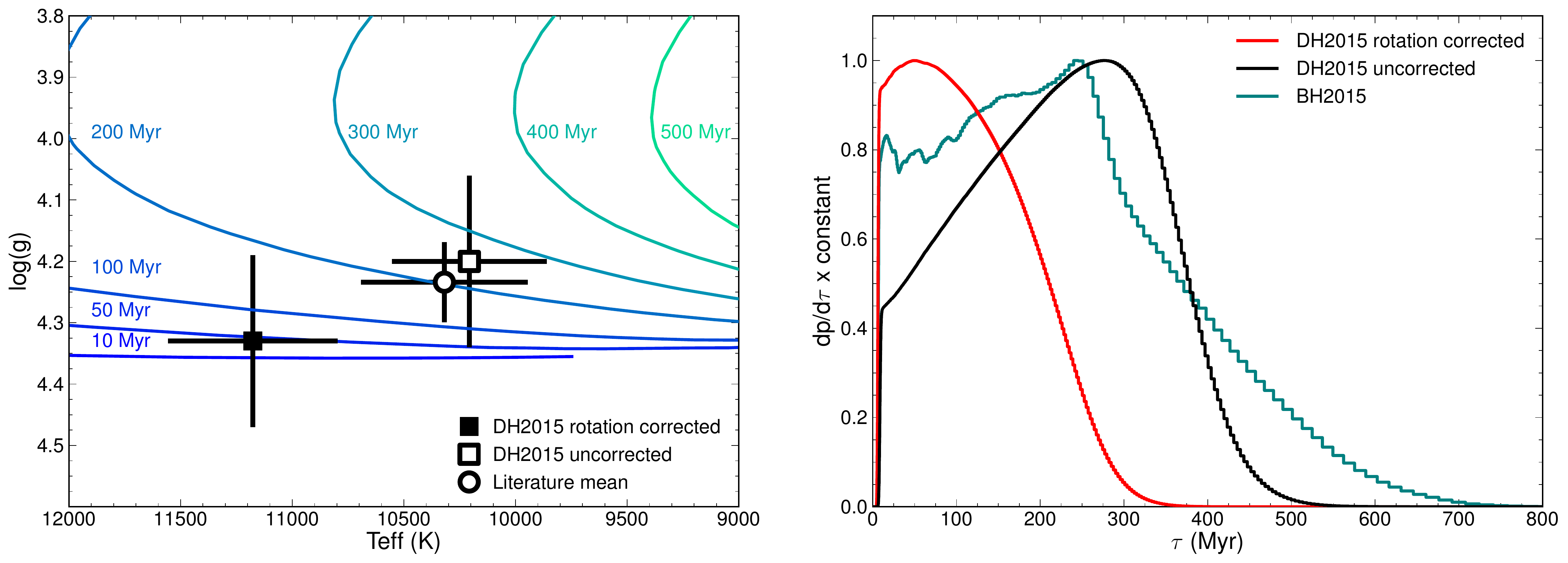}
\caption{Left: Determinations of the star's location in $T_\mathrm{eff}-\log{g}$ space. The solid curves are solar-metallicity PARSEC v1.1 isochrones \citep{Bressan2012}. The open circle represents the literature mean and standard deviation, including measurements from \cite{David2015, 2012MNRAS.427..343M, 2012A&A...537A.120Z, 2006A&A...458..293P, 1999A&AS..137..273G}. The open and filled squares represent the DH15 uncorrected and DH15 rotation-corrected values, respectively. The rotation-corrected values are clearly discrepant. Right: Posterior probability distributions in age in both the rotation-corrected (red) and uncorrected (black) cases from DH15. The teal curve is the age posterior from \cite{Brandt2015}. Taken collectively, the age of \thisstar\ is in the range of  50-400 Myr.}
\label{fig:age}
\end{figure*}

As mentioned earlier, direct interpolation in an H-R diagram or color-magnitude diagram can lead to biases in derived ages. Following \citet{David2015}, we thus used a Bayesian analysis of the star's location in $\log{T_\mathrm{eff}}-\log{g}$ space relative to solar metallicity ($Z=0.015$) PARSEC v1.1 evolutionary models \citep{Bressan2012}, yielding a 68\% age confidence interval of 120-360 Myr, which is consistent with the average age of A0 field dwarfs. We performed likelihood calculations on a 1000$\times$1000 grid from 1 Myr-10 Gyr in age, and 1-10 $M_\odot$ in mass. We used the Salpeter IMF \citep{Salpeter1955} as the prior on stellar mass and a uniform prior in linear age (i.e. constant star formation rate). The atmospheric parameters are determined from combination of $uvby\beta$ photometry and ATLAS9 models \citep{Castelli2004, Castelli2006}, and the details of both the atmospheric characterization and age determination are presented in \cite{David2015}, hereafter DH15.

Intermediate-mass stars are rapid rotators, and the effect of this rotation is to make the star appear cooler, more luminous, and hence older than a non-rotating star of the same mass. Additionally, rapid rotators spend a longer time on the main sequence than slow rotators. Recently, DH15 and \citet{Brandt2015}, hereafter BH15, have shown that the rotational effects on the inferred ages of intermediate-mass stars are substantial.\footnote{We note these authors account for rotation in different ways when determining ages for such stars. While DH15 apply the corrective formulae of \cite{Figueras1998} to the atmospheric parameters prior to age determination (assuming $v\sin{i} \approx v_\mathrm{rot}$), BH15 marginalize over projected rotational velocity.} 

Consequently, we explored two possible solutions for the age of \thisstar. The first solution is based on the rotation-corrected atmospheric parameters derived in DH15, $T_\mathrm{eff}= 11176 \pm 380$  K, $\log{g}=4.33 \pm 0.14$ dex (see Figure~\ref{fig:age}). At this position, direct interpolation yields an age of $43^{+82}_{-37}$ Myr, which is consistent with a previous estimate of 55 Myr \citep{1985A&AS...60...99W}\footnote{\citet{1985A&AS...60...99W} discusses age errors in a general sense but does not provide information for assigning errors to the A0 stars in that sample.}. The Bayesian age analysis in this case yields a median age of 110 Myr with 68\% and 95\% confidence intervals of 10-150, 10-250 Myr, respectively. 

The second, more likely, solution is based on the uncorrected atmospheric parameters from DH15, listed in Table~\ref{tbl-1}. This set of atmospheric parameters yields a median Bayesian age and 68\% confidence interval of $230^{+130}_{-110}$ Myr, with a corresponding interpolated age of $200^{+20}_{-160}$ Myr (where the uncertainties are determined from Monte Carlo error propagation). We consider this solution more likely for the following reasons: (1) the effective temperature is more consistent with modern spectral type scales which suggest $T_\mathrm{eff} \approx 9700$ K at A0 \citep{2004IAUS..224....1A, 2013ApJS..208....9P}, (2) both $T_\mathrm{eff}$ and $\log{g}$ in this case are consistent with the mean of previous determinations from the literature, (3) this older age is more consistent with the average age of A0 field dwarfs, (4) as noted in \S~\ref{sec:section2}, the star is unlikely to be associated with known young moving groups, (5) there is some evidence presented in DH15 that the procedure used to determine the rotation-corrected parameters is over-aggressive, (6) there are no other significant indicators of youth for \thisstar, (7) the older age is consistent with other modern estimates: 300 $\pm$ 51 Myr \citep{2012A&A...537A.120Z} and 138 $\pm$ 98 Myr \citep{1999A&AS..137..273G}, both from H-R diagram analyses.

Figure~\ref{fig:age} (left) demonstrates the position of \thisstar in $\log{T_\mathrm{eff}}-\log{g}$ space, relative to evolutionary models. The difficulty of age-dating intermediate-mass stars on and near the main sequence is evident from the typically large uncertainties in surface gravity (or equivalently, luminosity). The effect of including rotation is also illustrated by the significantly different atmospheric parameters obtained.

Figure~\ref{fig:age} (right) shows the posterior probability distribution function (PDF) in age for \thisstar, originally derived in DH15. Also depicted is a PDF derived from the similar Bayesian approach to isochrone age-dating of \cite{Brandt2015}. The BH15 PDF was generated assuming a Gaussian prior on [Fe/H], with $\mu$=-0.1 and $\sigma$=0.2 dex, consistent with the distribution observed for intermediate-mass stars in the solar neighborhood.\footnote{The BH15 PDF was downloaded from \texttt{http://bayesianstellarparameters.info/}} The BH15 PDF is broader than that of DH15 due to the fact that those authors marginalize over mass, metallicity, rotational velocity, and inclination, three parameters which can substantially affect the inferred ages of intermediate-mass dwarfs. In contrast, DH15 marginalize over mass only, while the typically minor differences in metallicity are implicitly accounted for in the atmospheric parameter uncertainties. For comparison, the BH15 PDF yields a median age of 220 Myr, with 68\% and 95\% confidence intervals of 10-290, 10-500 Myr, respectively.

All estimates suggest the age is $<$500 Myr, and notably the three most recent estimates suggest $\tau >$ 200 Myr, which is consistent with the average age of A0 field dwarfs. The range of published ages illustrates the difficulty of age-dating on and near the main sequence, particularly for intermediate-mass stars for which empirical age-dating methods are either non-existent or uncalibrated. We adopt the median Bayesian age and 68\% confidence interval as the final age and uncertainties for \thisstar, $\tau \approx 230^{+130}_{-110}$ Myr \citet{David2015}. However, the literature range can also be used to infer the loosest reasonable constraints on the system age: 50-400 Myr. For completeness, we consider this broad range of plausible ages when inferring the companion mass.

\begin{deluxetable*}{lcccccc}
\tabletypesize{\scriptsize}
\tablecaption{Astrometry and photometry of the low-mass companion to \thisstar\label{tbl-2}}
\tablewidth{0pt}
\tablehead{
\colhead{Data set }	&\colhead{Filter }	&\colhead{SNR}  &\colhead{$\Delta \alpha$ ($\arcsec$)}	 &\colhead{$\Delta \delta$ ($\arcsec$)}	&\colhead{m} 	&\colhead{M$^{\mathrm{a}}$}}
\startdata
NACO 2013 		&Lp	        &$\simeq 6.25$     &$0\farcs 333\pm 0\farcs 009$	 &$-0\farcs 806 \pm 0\farcs 009$	&$13.85\pm 0.25$ 	&$9.03\pm 0.26$ \\ 
NACO 2015		&Lp		&$\simeq 3.1$   	&$0\farcs 334\pm 0\farcs 015$	 &$-0\farcs 788 \pm 0\farcs 015$	&$13.63\pm 0.5$ 	&$8.5\pm 0.505$ \\ 
\enddata
\end{deluxetable*}

\section{Photometry and properties of the companion} 

\begin{figure}[!ht]
  \centering
\includegraphics[width=9cm]{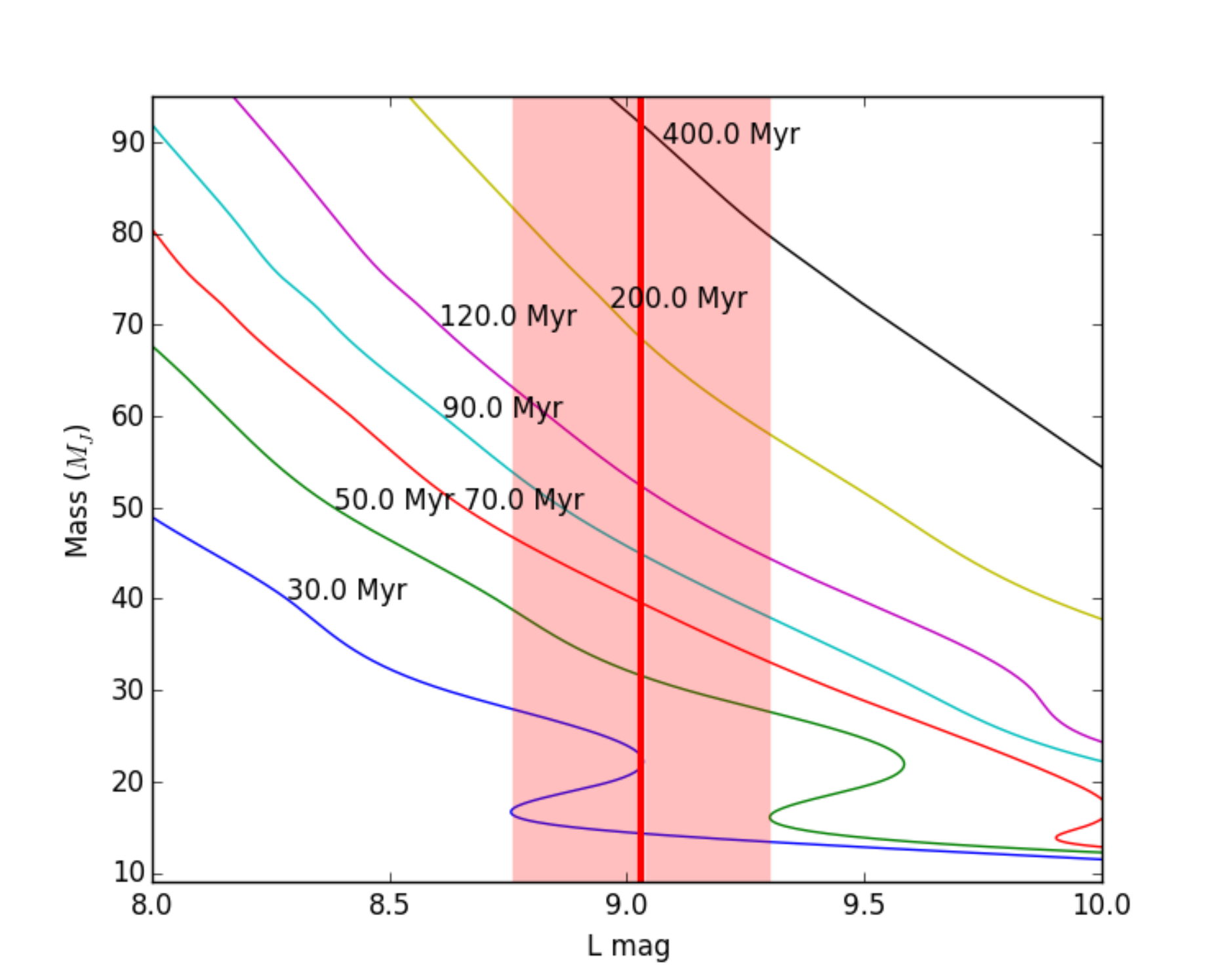}
  \caption{Mass ($M_J$) vs Luminosity (L mag) diagram for the evolutionary model BT-Settl. \thisstar B's L-band photometry and associated uncertainty is in shaded red.  \label{fig3}}
\end{figure}

The forward modeling MCMC approach to measure the companion photometry described above yields an absolute L magnitude of $9.03 \pm 0.26$ for the companion, including parallax/distance and L-band magnitude uncertainties. To infer the companion properties, we used the BT-Settl evolutionary model \citep{Allard2014} and our adopted conservative age range of 50-400 Myr. BT-Settl covers the range from solar-mass stars to the latest-type T and Y dwarfs, and reproduces the formation of clouds and in particular their clearing at the L/T transition. We derived a mass range of $15-90\ M_J$ (Figure \ref{fig3}), and effective temperature between 1900K and 2700K, placing \thisstar B in the L-dwarf regime. 

\section{Discussion}
No far-infrared observations of \thisstar are listed in the data archives for the Spitzer or Herschel space telescopes, so the properties of its debris disk can only be estimated from the WISE survey data. The AllWISE survey magnitudes are 6.04$\pm$0.05, 5.97$\pm$0.04, 6.06$\pm$0.02, and 5.50 $\pm$0.03 - corresponding to 3.6, 4.5, 11.8, and 22.0 $\mu$m respectively. An excess of 0.56 $\pm 0.06$ mag is detected at 22 $\mu$m, but no excess is evident at shorter wavelengths. Using a 3 $\sigma$ upper limit to the 11.8 $\mu$m excess (0.05 mag), and assuming blackbody emission, the upper limit to the disk dust temperature is 168 K. For grain sizes of a few microns, this corresponds to a disk inner edge of $r > 10$ AU.  

While the WISE data provide no constraint on the disk outer radius, \thisstar B appears to be massive enough to gravitationally clear any disk material in its vicinity.  It is unlikely that any disk material located beyond the companion's $>$80 AU orbital separation would be warm enough ($\sim$ 120 K) to emit significantly at 22 $\mu$m.  The most likely scenario is therefore that \thisstar B orbits exterior to a warm dust belt that is the source of the WISE excess.  An exterior cold dust belt beyond \thisstar B orbiting radius may also exist but would require ALMA observations to detect.  

The \thisstar system shares several features with the $\kappa$ Andromedae system \citep{Carson13, 2013ApJ...779..153H}. Like $\kappa$ Andromedae, \thisstar is a late-B/early-A type star with a mass $\sim$3 $M_\odot$, and a poorly constrained age. Further, the companions in both systems are ``brown dwarf desert'' objects with masses in the range of $\sim$15-80 M$_{Jup}$, corresponding to mass ratios of $\sim$1\%. Such ``extreme mass ratio systems'' \citep[e.g.][]{Hinkley2015}, are particularly important for constraining the formation mechanisms of stars in this mass range. Indeed, several works \citep[e.g.][]{dcb04,kbp07} have suggested that the multiplicity of intermediate-mass stars may serve as a reservoir for the conserved initial angular momentum in the protostellar cloud.  

Lastly, we note that no other companion more massive or similar to \thisstar B is detected around \thisstar from the effective inner working angle of $\simeq 0".3$ (projected separation of $\simeq 30$ AU) to the outer edge of the effective field of view of $\simeq 12"$ ($\simeq 1000$ AU).

\section{Conclusion}
This paper presented the detection of a substellar companion orbiting at a projected separation of $\simeq 80$ AU around \thisstar, a disk-bearing A0V star. The characterization of the companion is made difficult by the uncertain age determination for field A stars. Spectroscopic follow-up using medium resolution slit spectroscopy is the next step to further characterize the low-mass object around \thisstar \citep{Hinkley2015b}. Indeed spectral indices to quantitatively measure the strength of the FeH, VO, KI, spectral features have been demonstrated to be robust age markers, but require medium resolution spectroscopy ($R>300$).

\acknowledgments
The first author is grateful to Prof. Lynne Hillenbrand (Caltech) and Ben Zuckerman (UCLA) for their advice and comments. T.J.D is supported by the National Science Foundation Graduate Research Fellowship under grant No. DGE1144469. M.B. is supported by a NASA Space Technology Research fellowship, grant No. NNX13AN42H. This publication makes use of data products from the Wide-field Infrared Survey Explorer, which is a joint project of the University of California, Los Angeles, and the Jet Propulsion Laboratory/California Institute of Technology, funded by the National Aeronautics and Space Administration.




{\it Facilities:} \facility{Very Large Telescope}



\end{document}